\documentclass[sigplan,screen,authorversion,nonacm]{acmart}

\AtBeginDocument{%
  \providecommand\BibTeX{{%
    \normalfont B\kern-0.5em{\scshape i\kern-0.25em b}\kern-0.8em\TeX}}}

\setcopyright{acmlicensed}
\acmPrice{15.00}
\acmDOI{10.1145/3520313.3534660}
\acmYear{2022}
\copyrightyear{2022}
\acmSubmissionID{pldiws22soapmain-id8-p}
\acmISBN{978-1-4503-9274-7/22/06}
\acmConference[SOAP '22]{Proceedings of the 11th ACM SIGPLAN International Workshop on the State Of the Art in Program Analysis}{June 14, 2022}{San Diego, CA, USA}
\acmBooktitle{Proceedings of the 11th ACM SIGPLAN International Workshop on the State Of the Art in Program Analysis (SOAP '22), June 14, 2022, San Diego, CA, USA}

\usepackage{xcolor}

\definecolor{commentsColor}{rgb}{0.497495, 0.497587, 0.497464}
\definecolor{keywordsColor}{rgb}{0.000000, 0.000000, 0.635294}
\definecolor{stringColor}{rgb}{0.558215, 0.000000, 0.135316}
\colorlet{darkblue}{blue!30!black}
\colorlet{darkgreen}{green!30!black}

\usepackage{amsmath}
\usepackage{stmaryrd}

\usepackage{paralist}

\usepackage{multirow}
\usepackage{multicol}

\usepackage{microtype}

\usepackage{xspace}
\usepackage{xparse}
\usepackage{listings}
\lstdefinelanguage{ocaml}
{%
        keywords=[2]{assert,else,entry,failwith,fun,if,in,let,match,of,storage,then,type,with,\$storage, \$sender, \$amount},%
        keywordstyle=[2]\color{darkblue},
        sensitive=true, 
        morecomment=[l]{//}, 
        morecomment=[s]{(*}{*)}, 
        morestring=[b]", 
        keywords=[3]{bool, byte, int, key, key\_hash, lambda, list, map,
        big\_map, nat, option, or, pair, set, signature, string, bytes, mutez,
        timestamp, unit, operation, address, ticket},
        keywordstyle=[3]\color{darkgreen},
}
\lstdefinelanguage{Michelson}
{%
        keywords=[2]{ABS, ADD, ADDRESS, AMOUNT, AND, BALANCE, BLAKE2B, CAR, CAST,
                CDR, CHECK\_SIGNATURE, COMPARE, CONCAT, CONS, CONTRACT, CREATE\_ACCOUNT,
                CREATE\_CONTRACT, DIP, DROP, DUG, DIG, DUP, EDIV, EMPTY\_MAP, EMPTY\_SET, EQ, EXEC,
                FAILWITH, GE, GET, GT, HASH\_KEY, IF, IF\_CONS, IF\_LEFT, IF\_NONE,
                IMPLICIT\_ACCOUNT, INT, ISNAT, ITER, LAMBDA, LE, LEFT, LOOP, LOOP\_LEFT, LSL,
                LSR, LT, MAP, MEM, MUL, NEG, NEQ, NIL, NONE, NOT, NOW, OR, PACK, PAIR, PUSH,
                RENAME, RIGHT, SELF, SENDER, SET\_DELEGATE, SHA256, SHA512, SIZE, SLICE, SOME,
        SOURCE, STEPS\_TO\_QUOTA, SUB, SWAP, TRANSFER\_TOKENS, UNIT, UNPACK,
        UPDATE, XOR, UNPAIR, ASSERT_CMPEQ, FAIL},%
        keywordstyle=[2]\color{darkblue},
        sensitive=true, 
        morecomment=[l][\color{darkgreen}]{\#}, 
        morecomment=[s]{(*}{*)}, 
        commentstyle=\color{darkgray}\ttfamily,
        morestring=[b]", 
        keywords=[3]{bool, contract, int, key, key\_hash, lambda, list, map,
        big\_map, nat, option, pair, set, signature, string, bytes, mutez,
        timestamp, unit, operation, address, ticket, union},
        keywordstyle=[3]\color{darkblue},
}
\lstdefinestyle{mystyle}{%
        backgroundcolor=\color{white},   
        basicstyle=\linespread{0.7}\scriptsize\ttfamily,        
        breakatwhitespace=false,         
        breaklines=true,                 
        captionpos=b,                    
        columns=fixed,                   
        commentstyle=\color{commentsColor}\textit,    
        escapeinside={\%*}{*)},          
        extendedchars=true,              
        float,
        floatplacement=tbp,
        keepspaces=true,                 
        keywordstyle=\color{keywordsColor}\bfseries,       
        language=Michelson,                      
        numbers=left,                    
        numbersep=5pt,                   
        numberstyle=\tiny\color{commentsColor}, 
        otherkeywords={},           
        rulecolor=\color{black},         
        showspaces=false,                
        showstringspaces=false,          
        showtabs=false,                  
        stepnumber=1,                    
        stringstyle=\color{stringColor}, 
        tabsize=2,	                 
        texcl=true
}
\NewDocumentCommand{\micheltype}{v}{\lstinline[language=michelson,basicstyle=\small\ttfamily]{#1}}
\NewDocumentCommand{\micheldomain}{v}{\lstinline[language=michelson,basicstyle=\small\ttfamily]{#1}}
\NewDocumentCommand{\michelinstr}{v}{\lstinline[language=michelson,basicstyle=\small\ttfamily]{#1}}

\newcommand{\ie}{\textit{i}.\textit{e}. }

\usepackage{hyperref}
\newcommand{\fref}[1]{%
  \hyperref[#1]{Fig.~\ref*{#1}}\xspace
}
\newcommand{\tref}[1]{%
  \hyperref[#1]{Table~\ref*{#1}}\xspace
}
\newcommand{\sref}[1]{%
  \hyperref[#1]{Sect.~\ref*{#1}}\xspace
}
\newcommand{\sectionref}[1]{%
  \hyperref[#1]{Section~\ref*{#1}}\xspace
}
\newcommand{\appendixref}[1]{%
  \hyperref[#1]{Appendix~\ref*{#1}}\xspace
}

\usepackage[skip=2pt]{caption}

\AtBeginDocument{%
  \addtolength{\belowcaptionskip}{-0.4\baselineskip}%
  \addtolength\abovedisplayskip{-0.4\baselineskip}%
  \addtolength\belowdisplayskip{-0.4\baselineskip}%
}

\begin{document}

\title{Abstract Interpretation of Michelson Smart-Contracts}\thanks{This work is
        partially supported by the European Research Council under Consolidator
Grant Agreement 681393 — MOPSA}

\author{Guillaume Bau}
\affiliation{%
        \institution{Sorbonne Université, CNRS, LIP6, Nomadic Labs}
        \city{Paris}
        \country{France}
}
\email{guillaume.bau@nomadic-labs.com}

\author{Antoine Miné}
\affiliation{
        \institution{Sorbonne Université, CNRS, LIP6}
        \city{Paris}
        \country{France}
}
\email{antoine.mine@lip6.fr}

\author{Vincent Botbol}
\affiliation{%
        \institution{Nomadic Labs}
        \city{Paris}
        \country{France}
}
\email{vincent.botbol@nomadic-labs.com}

\author{Mehdi Bouaziz}
\affiliation{%
        \institution{Nomadic Labs}
        \city{Paris}
        \country{France}
}
\email{mehdi.bouaziz@nomadic-labs.com}

\renewcommand{\shortauthors}{Guillaume Bau, Antoine Miné, Vincent Botbol, and
Mehdi Bouaziz.}

\newcommand{\Michelson}{\emph{Michelson}\xspace}

\begin{abstract}
        Static analysis of smart-contracts is becoming more widespread
        on blockchain platforms. Analyzers rely on techniques
        like symbolic execution or model checking, but few of them can
        provide strong soundness properties and guarantee the analysis
        termination at the same time. As smart-contracts often
        manipulate economic assets, proving
        numerical properties beyond the absence of runtime errors
        is also desirable. Smart-contract
        execution models differ considerably from mainstream
        programming languages and vary from one blockchain to another,
        making state-of-the-art analyses hard to adapt. For instance,
        smart-contract calls may modify a persistent storage
        impacting subsequent calls. This makes it difficult for tools to
        infer invariants 
        required to formally ensure the absence of exploitable vulnerabilities.

        The \Michelson smart-contract language, used in the Tezos
        block\-chain, is strongly typed, stack-based, and has a strict
        execution model leaving few opportunities
        for implicit runtime errors. We present a work in progress static analyzer for \Michelson
        based on Abstract
        Interpretation and implemented within MOPSA, a modular
        static analyzer.
        Our tool supports the \Michelson semantic features, including inner calls to external
        contracts. It can prove the absence of runtime errors and infer
        invariants on the persistent storage over an unbounded number
        of calls. It is also being extended to prove high-level
        numerical and security properties.
\end{abstract}




\maketitle

\section{Introduction}
\subsection{Blockchains and Smart-Contracts}
Blockchains are distributed immutable ledgers organized in peer to peer
networks, allowing participants to securely transfer tokens
without a central authority.
Some blockchains allow programmable transactions in the form
of computer programs. These \emph{smart-contracts} define
complex transactions between blockchain participants and maintain
a persistent state across runs. They can be viewed as a novel way
for multiple users to securely exchange, build value sharing or distribution
applications without a trusted third-party. Applications include auction
sales, decentralized exchanges, collective organizations, investment funds, etc.

Smart-contracts are relatively small, and not resource intensive as they have to be executed on all blockchain network nodes.
However, compared to usual programming languages, they have an unconventional
execution model tightly tied to the blockchain implementation, thus are
non-intuitive to program.
Compounded with the inability to update smart-contracts on the (immutable) blockchain, and applications manipulating large sums of money, this leads to costly errors.
Notable vulnerability examples include: a reentrancy issue in
\emph{The DAO}~\cite{del2016dao},
a smart-contract implementing a venture capital
fund, that allowed a user to steal \$60 million; the \emph{Parity} wallet
bug~\cite{palladino2017parity}, which froze \$150 million by allowing
unauthenticated users to call restricted functions; and the
Proof-of-Weak-Hands attack, that allowed attackers to steal \$800,000
overnight~\cite{powh:1}, and \$2.3 million then after, abusing an integer
overflow. Thereby, there is a high motivation to statically detect potential
misbehavior or prove their absence when possible.

\subsection{Motivating Example}
\begin{figure}[t]
  \begin{lstlisting}[language=ocaml,xleftmargin=10pt]
storage : (address, mutez) map
entry deposit () {
   let owned = match Map.get $storage $sender with
     | None -> 0
     | Some v -> v in
   (Map.add $sender (owned + $amount) $storage, [])
}
entry withdraw (asked : mutez, dest : address) {
   assert ($amount == 0);
// Fix to ensure proper access control:
// if dest != \$sender then failwith "unauthorized";
   let owned = match Map.get $storage dest with
     | None -> failwith "empty account"
     | Some v -> v in
   if asked > owned then
     failwith "not enough tokens";
   (Map.add dest (owned - asked) $storage,
    [transfer $sender asked])
}\end{lstlisting}
  \caption{A smart-contract with incorrect authentication}  \label{fig:motiv}
\end{figure}

We focus on the verification by static analysis of
smart-contracts for the Tezos blockchain programmed in the Michelson
\cite{michelson:1, michelson:2} language.
Our goal is to analyze \emph{Dexter}~\cite{dexter:1}, an important smart-contract
implementing a decentralized exchange with alternate blockchain currencies
(\emph{bitcoins}, \emph{usdtz}). Its initial version featured a vulnerability~\cite{dexter:bug}
allowing an attacker to steal parts of the contract funds.
As this is a work in progress, we report on a preliminary analysis of a
simplified version only.

Consider the contract in \fref{fig:motiv} inspired from Dexter
and written in an ML-style pseudo-code.
It implements a simple wallet, allowing users to deposit to or withdraw from
their personal account some amount in \micheltype{mutez}
(the currency on the Tezos blockchain).
A \emph{map} keeps track of user accounts: it maps users, identified by their blockchain \emph{address},
to the deposited amount.
The map is kept on the blockchain, in a so-called \emph{storage}, updated after
each transaction.
An execution of the smart-contract starts with the following variables:
\begin{compactitem}[$\bullet$]
\item \michelinstr{$storage} is the storage value currently on the blockchain.
\item \michelinstr{$sender} is the address of the initiator of the contract call
  (a user, or another smart-contract).
\item \michelinstr{$amount} is the amount of mutez transferred to the contract.
  Every call is a transfer, possibly with a 0 amount. 
\end{compactitem}
A contract can define several independently callable entry points. They allow
splitting functionalities sharing the same storage.
In our case:
\begin{compactitem}[$\bullet$]
\item \emph{deposit} allows a user to deposit an amount. The
  \michelinstr{$amount} sent to the contract is actually
  recorded to belong to the user by updating his balance in
  the map (\texttt{Map.add}, line~6).
\item \emph{withdraw} allows a user to transfer back some
  amount from the contract, unless his account in the map
  does not hold sufficient funds (\texttt{if asked > owned}, line~15).
  The map is updated (\texttt{Map.add}, line~17) and a transfer back
  to the sender is generated (\texttt{transfer}, line~18).
\end{compactitem}
Michelson features a purely functional execution model: the new value
of the storage as well as any effect (e.g., additional transfers) are
provided in the return value of the call.

This example actually contains a logic error: it allows a user to
transfer to himself \micheltype{mutez} that were owned by someone else.
Indeed, the \emph{dest} parameter used as key in the map in \texttt{withdraw}
is controlled by the user who calls the contract.
A fix is provided in a comment line~11: it ensures that \texttt{dest} equals
the caller \michelinstr{$sender}.
In this example, we want to verify the high-level property stating that:
$$(key = \$sender) \lor (new\_value \geq old\_value)$$
whenever the map is updated on \texttt{key} from \texttt{old\_value} to
\texttt{new\_value}, i.e., a user can only add funds to another user's account
and can subtract funds only from his own account.

\subsection{Related Work}
A number of tools have been designed to help smart-contract developers
catch bugs or vulnerabilities. Most of them target the \emph{Ethereum}
platform. This includes symbolic execution tools,
like Maian~\cite{nikolic2018finding}, Manticore~\cite{mossberg2019manticore},
Oyente~\cite{luu2016making}, Zeus~\cite{kalra2018zeus},
Securify~\cite{tsankov2018securify}, Mythril~\cite{mueller2020introducing}.
Some tools rely on exhaustive state exploration, via model checking or
SMT solving, sometimes leading to a slow analysis~\cite{aryal2021comparison},
timeouts~\cite{ren2021:1}, or lack of results~\cite{aryal2021comparison}.
\cite{kalra2018zeus} relies on Abstract Interpretation for a
preliminary analysis, and uses an SMT solver to check properties
on inferred invariants.
\cite{ghaleb2020effective,durieux2020empirical,clairvoyance:1,dia:1}
report that many existing tools fail to detect some issues, i.e., report false
negatives.
\cite{ghaleb2020effective,schneidewind2020ethor,clairvoyance:1}
affirm that some tools can fail to prove properties because their analyses are
unsound.

Some tools focus on low-level properties affecting the popular Ethereum
platform, like reentrancy issues~\cite{rodler2018sereum,liu2018reguard} or
overflows~\cite{easyflow:1}.
By contrast, Michelson~\cite{michelson:1, michelson:2}, the language of the Tezos blockchain,
which is the focus of our work, has fewer opportunities for runtime errors and
a stricter execution model eliminating reentrancy issues; low-level properties
are less of an interest.
Checking higher-level properties is feasible using proof assistants,
but requires a large effort to
prove even simple properties. Mi-cho-Coq~\cite{michocoq:1} provides a
Michelson Coq embedding allowing the certification of smart-contract properties, but
requires manual developments of proofs, and small changes in a contract
require new proof developments. The Micse project~\cite{micse:1} allows for
automated static analysis, using the Z3 SMT solver. The
Tezla~\cite{reis2020tezla} project allows translating the Michelson instructions
into a suitable intermediate representation for dataflow analysis.

\subsection{The MOPSA Static Analyzer}

\emph{MOPSA}~\cite{mine-VSTTE19} is a modular and extensible static analyzer
based on Abstract Interpretation~\cite{cousot1977abstract}.
It features a C analyzer detecting runtime errors and invalid preconditions
when calling the C library, as well as Python type, value, and uncaught
exception analyses.
Its modular design allows sharing and reusing abstract domains across
multiple analyses. Its \emph{AST} structure can be extended
to support novel languages, keeping a high-level representation
without static translation.

MOPSA strongly relies on domain cooperation.
An analysis is defined as a combination of small domain modules, including
value abstractions and syntax iterators that can
be plugged in or out, depending on the target language and properties.
It provides a common set of domains to build value analyses
with intervals, relational domains like
octagons \cite{mine-HOSC06} or polyhedra \cite{ch:popl78}
to infer linear relations, recency abstraction \cite{recency} for
memory blocks, etc.
In addition to reductions, domains cooperate through expression rewriting.
For instance, a domain handles C arrays by rewriting array accesses
dynamically as accesses into scalar variables representing array cells.
Expression rewriting helps writing small, independent, and reusable
domains, that rely only on the manipulation of variables the state
of which is managed by other, lower-level domains.

\subsection{Contribution}

We have developed an analysis of Michelson programs \cite{michelson:1, michelson:2}
based on Abstract Interpretation. This analysis is built on MOPSA
\cite{mine-VSTTE19}. It reuses domains provided by MOPSA and provides
novel domains to support the semantics of the Michelson language.
This includes support for specific Michelson, ML-like types, such as pairs,
unions, sets, maps, etc. as well as iterators to handle the execution model for
contracts on the Tezos platform, including contract interactions.
Our tool can currently statically detect runtime errors
like \emph{overflows}, \emph{shift overflows}, and Michelson contracts
always terminating in a failure state.
We also demonstrate its potential to prove higher-level correctness
properties on the example contract from \fref{fig:motiv}.
By using abstract interpretation, our analysis is sound and efficient,
but can raise false alarms.

\sectionref{sec:simpleanalysis} presents a basic set of
abstract domains that are sufficient to cover the complete semantic
of Michelson instructions and achieve an initial, sound, low-precision
analysis;
\sref{sec:executionmodel} presents
the support for the Tezos transaction execution model, including
contracts calling external contracts and inferring invariants on unbounded
sequences of calls to contracts;
\sref{sec:higher} presents more involved abstractions necessary
to prove the correctness of \fref{fig:motiv};
\sref{sec:experiments} presents our experimental evaluation;
\sref{sec:conclusion} concludes.

\section{Michelson Value Analysis}\label{sec:simpleanalysis}

For convenience, \fref{fig:motiv} presented a
smart-contract example in a high-level, ML-like syntax.
Michelson \cite{michelson:1}, the language actually executed
on the Tezos blockchain, is a high-level \emph{stack-based} language
that takes inspiration from Forth~\cite{forth} or Joy~\cite{joy:1},
while including many aspects from functional languages:
strong static typing, immutable values, anonymous functions,
algebraic data-types, functional \emph{list}, \emph{set},
and \emph{map} types.
This section presents the abstract domains added to MOPSA
to handle the stack, data-types, and instructions.

\subsection{The Michelson Language}\label{sec:detailedexample}

\begin{figure}[t]
\begin{lstlisting}[xleftmargin=10pt]
storage nat;
parameter nat;
code {  UNPAIR;
        ADD;
        NIL operation;
        PAIR; }
\end{lstlisting}
\caption{Simple Michelson smart-contract}\label{fig:sample1}
\end{figure}

To simplify, we consider here a much simpler
contract, in \fref{fig:sample1}, that performs an
addition into an accumulator stored on the blockchain.

In Michelson, there are no explicit variables.
All values are stored on a stack, implicitly manipulated through
dedicated instructions such as \michelinstr{PUSH}, \michelinstr{DROP},
\michelinstr{DUP}, and using operator instructions (e.g., \michelinstr{ADD}
to perform an addition) replacing arguments at the top
of the stack with the operator result.

When the contract execution starts, the stack contains a
single element: a pair containing the value of the parameter
it has been called with and the value stored on the blockchain
for the contract.
When the execution ends, the contract should leave on
the stack a single value: a pair containing a list of
operations to perform after the contract execution (such as
calling other contracts, or performing a transfer) and
the new value to be stored on the blockchain.
Operations will be discussed in details in \sref{sec:executionmodel}.
For now, the operation list output by a contract will be empty.
The initial value of the storage is specified when the contract
is deployed on the blockchain.
Subsequent executions of the contract update the storage
value.

As the language is statically typed, a contract declares the
type of its storage and parameter.
This corresponds to lines~1--2 in \fref{fig:sample1}.
In the example, both the storage and
parameter have type \micheltype{nat} (i.e., a non-negative integer),
but more complex data structures can be used. For instance,
\fref{fig:motiv} uses a map as storage and its
parameter is a union to model different possible entry points.

When executed, the code from \fref{fig:sample1} proceeds as follows:
\begin{compactitem}[$\bullet$]
\item \michelinstr{UNPAIR} pops the topmost (and only) element from the stack: a pair
with the storage and the parameter, which are pushed as the first and second
items on the stack;
\item \michelinstr{ADD}, pops two elements, adds them, and pushes the result;
\item \michelinstr{NIL operation} builds an empty list of operations, and\\ pushes
it on the stack;
\item \michelinstr{PAIR} pops the addition result and the empty list, and pushes a pair,
  resulting in a stack with a single pair element.
\end{compactitem}
Thus, each call to the contract will simply add the integer passed
as parameter to the integer stored on the blockchain.

\subsection{Dynamic Translation into Variables}

MOPSA models the memory as a map from variables to values and supports
instructions, such as assignments and tests, involving expressions
over variables.
This is a common assumption for abstract interpreters as well as
domain libraries (such as APRON \cite{apron:1},
used in MOPSA) and especially useful for relational analyses.
One possibility to handle stack-based languages
is to translate them to variable-based environments and
expressions beforehand, in a pre-processing phase, as performed
for instance in the Sawja framework \cite{sawja} for Java bytecode
as well as Tezla \cite{reis2020tezla} for Michelson.

Instead of a static translation, we extended MOPSA's AST
with a native support for Michelson instructions and relied
on the ability for domains to rewrite statements and expressions as
part of the abstract execution.
We developed a domain that introduces variables to represent stack positions
and translates Michelson instructions into assignments on-demand.
Non-scalar data-types, such as pairs, give rise to several variables per
stack position, as detailed in \sref{sec:types}.
A dynamic translation can potentially use information about the current
precondition to optimize the translated instructions \cite{mine-VSTTE19},
although this is not currently the case for Michelson.

\subsection{Michelson Data-Types}
\label{sec:types}


Michelson supports several integer kinds: arbitrary precision integers
(\micheltype{int}), natural integers (\micheltype{nat}), dates,
and unsigned 63-bit integers (\micheltype{mutez}).
Some operations, such as overflows on \micheltype{mutez}, as well as
shift overflows, are checked runtime errors that halt the contract
execution.
A specific domain in MOPSA handles these types, checking all possible
runtime errors and representing their possible values in standard
numeric domains such as intervals and polyhedra.


Michelson supports simple algebraic types \emph{à la ML} through pairs
(\michelinstr{(a,b)}), option types (\michelinstr{Some a} or \michelinstr{None}), and
tagged unions with two variants (\michelinstr{Left a} or \michelinstr{Right b}).
The type of \michelinstr{a} and \michelinstr{b} is arbitrary, and algebraic
types can be nested.
MOPSA features domains to handle these types.
They create and manage additional variables for each component of
a pair, an option, or a sum, delegate the abstraction of
their value to the domain of the components' type, and translate
operations on algebraic types (such as \michelinstr{PAIR},
\michelinstr{CAR}, etc.) into operations on component variables.
Domains handling scalar values, such as numeric domains, ultimately
work on environments mixing components from different algebraic
values, making it possible to infer relations between values
that appear inside pairs or options.
This technique is similar to that of Bautista et al. \cite{nsad:2020}, but
we support recursive types and do not partition with respect to
which variant is used by each variable.


Michelson has a native support for immutable containers: lists, sets,
and maps.
We propose simple, general-purpose, and efficient, but coarse abstractions to handle
them.
Lists are abstracted using a summary variable to represent the union of all
list elements, and a numeric variable representing its size.
Like algebraic types, list elements can have arbitrary type.
List operations are translated into operations on the variables
(e.g.,  weak updates of summary variables, size incrementation) and delegated
to the domain appropriate for the type. List iteration \michelinstr{ITER} is handled, as usual in
abstract interpretation, using a fixpoint.
Sets are abstracted similarly to lists with slight adjustments as they cannot
contain duplicate elements.
Maps are abstracted using a summary variable to represent keys and
a summary variable to represent values.
A more involved, property-specific abstraction of maps will be discussed in \sref{sec:smap}.

\subsection{Addresses}\label{sec:addresses}
Michelson has a domain-specific type for addresses, representing
participants on the blockchain: either users (identified by a
public key) or smart-contracts (identified by a hash).
Some addresses play a special role during contract execution, and can
be accessed using dedicated instructions.
As detailed in \sref{sec:executionmodel}, a contract execution
can be triggered by the execution of another contract.
\michelinstr{SOURCE} represents the user at the origin of
a chain of calls, while \michelinstr{SENDER} is the immediate
caller of the contract.
These variables play an important role in access control and thus
the security of contracts, as demonstrated by the fix proposed
line~11 of \fref{fig:motiv} for our incorrect wallet implementation.

We use a reduced product of two domains for addresses:
a powerset of address constants -- useful to
precisely handle addresses hard-coded in a contract -- and
a domain that maintains whether the address equals \michelinstr{$sender} or not.
It is useful to handle precisely access control by comparison with
\michelinstr{$sender}, which is not a literal constant.

\section{Execution Model and Analysis}\label{sec:executionmodel}

The previous section presented domains sufficient to handle the
execution of arbitrary Michelson code on an input stack.
In this section, we take into account the execution in its context
on the blockchain.
A contract can be executed multiple times, making its storage evolve
during time.
Additionally, one execution can trigger additional contract executions
through the operation list it returns.

\subsection{Execution Context}

Once deployed (originated) on the blockchain, smart-contracts are available for
any user to call.
A call must provide a parameter as well as an entry point for the contract.
As different entry points execute very different code, MOPSA performs a case
analysis: for each entry point, the contract is analyzed on an initial stack
for this entry point with a corresponding abstract parameter value modeling
any possible actual value in the parameter type; the results are then joined
after execution.
The execution context also sets up special variables, such as \michelinstr{$sender},
modeling the contract caller and initialized with a symbolic value in the
address domain (\sref{sec:addresses}).

An analysis of the contract on an initial, empty storage would only model the
very first execution of the contract, which is not sound.
For instance, in \fref{fig:motiv}, an empty storage means
that the \emph{withdraw} entry point always fails.
Alternatively, starting with an abstract storage representing all possible
concrete values in its type could be imprecise.
\sectionref{sec:multcalls} will propose another solution where a sound
abstraction of the storage is inferred through fixpoint computation.

\subsection{Operation List}
In addition to the updated storage, Michelson contracts can return
a list of operations to execute after they finish.
These operations can be some calls to smart-contracts, which entails
executing these contracts with the updated storage.
These can, in turn, append new operations to the operation list.
The list is traversed in depth-first order until there are no
more operations to execute.
Note that a contract cannot call another contract in the middle of its
execution and expect a return value; moreover, the execution of the operation
list is atomic: a runtime error at any point reverts all modifications to
the storage of the contracts involved.
This unusual execution model makes reentrancy bugs, such as the one
plaguing as \emph{The DAO}~\cite{del2016dao}, less likely on Tezos.

MOPSA has partial support for this model.
We do compute the operation list and iterate contract execution in a
fixpoint, using updated storage and inferred entry points and arguments,
as mandated.
This includes the cases where a contract calls itself, or another contract.
However, our coarse abstraction of lists using summary variables
(\sref{sec:types}) makes the analysis impractical when a contract calls
more than one contract.
It should be addressed in future work.

\subsection{Multiple Calls Analysis}
\label{sec:multcalls}
Analyzing a unique call to a smart-contract provides
some insights on the possible runtime errors, but it does not take
into account all possible executions over its whole lifetime.
We developed an analysis to over-approximate an infinite number of calls to a
smart-contract, from different callers and to multiple entry points.

Let \emph{Addr} be the set of all addresses, \emph{Entrypoints} the set of entry points for the
contract and $P_e$ the semantic function computing the new storage
after executing entry point $e$ of contract \emph{P}. The next storage $S_{i+1}$ of
the contract as a function of its current storage $S_i$
(assuming, for the simplicity, an empty list of operations) is:
$$\mathrm{S}_{i+1} = \exists \mathrm{addr} \in \mathrm{Addr}, \exists \mathrm{e} \in \mathrm{Entrypoints}, \mathrm{call(P_e, addr, S_{i})}$$

Using the (classic) technique of iterations with widening, with \$sender being
an abstract value of our address reduced product from \ref{sec:addresses}, our
analysis computes an abstraction of the fixpoint:
$$\mathit{lfp}_{S_0} \left( \lambda S: \bigsqcup_{\forall e \in Entrypoints}
\mathrm{call(P_e, \$sender, S)} \right)$$
which models arbitrary sequences of executions of the contract from the initial
storage $S_0$ .
It thus outputs all possible runtime errors.
It also returns an invariant over storage values, which could be inspected by
the user for additional insight on the behavior of the smart-contract.
On the example of \fref{fig:motiv}, we discover that the \emph{deposit}
entry point will fill an initially empty map and allow some \emph{user} to
call the \emph{withdraw} entry point without entering the
\emph{failure} state.
This is not sufficient yet to prove our property of interest, that
``only owners can decrease the amount of tokens in the map.''

\section{High-Level Domains}
\label{sec:higher}

We now present additional abstract domains, bringing more precision
necessary to analyze our motivating example.

\subsection{Symbolic Expressions}
\begin{figure}[tb]
  \begin{lstlisting}[xleftmargin=10pt]
// assuming a stack containing values  x :: y;
DIP { DUP } // x :: y :: y
DUP;        // x :: x :: y :: y
DUG 2;      // x :: y :: x :: y
COMPARE;    // pops 2 items, push -1, 0 or 1
EQ;         // boolean test if -1, 0 or 1 equals to 0
IF          // pops boolean and branches accordingly
   { } // x = y
   { } // x != y
\end{lstlisting}
  \caption{Comparison in Michelson}\label{fig:compare}
\end{figure}

In Michelson, there is no direct comparison operator.
Consider the example in \fref{fig:compare} that executes different branches
when the topmost stack elements \texttt{x} and \texttt{y} are equal, and when
they are different.
The \michelinstr{COMPARE} polymorphic instruction pushes $-1$ (resp. $0$, $1$)
on the stack when one operand is smaller than (resp. equal to, greater than)
the other.
Then, an integer operation such as \michelinstr{EQ} compares the result to $0$
and pushes a boolean on the stack, which is consumed by \michelinstr{IF}.
To be precise, an analysis must track this sequence of instructions.
Using the domains presented in \sref{sec:simpleanalysis}, our analysis is only
able to infer that \emph{true} or \emph{false} is pushed on the stack and
immediately consumed, inferring no information on the
topmost stack values \texttt{x} and \texttt{y}  inside the branches.

As an alternative to developing a complex relational domain, we implemented
the symbolic constant abstract domain proposed in \cite{mine-VMCAI06}.
This domain assigns to each variable a value $v$ from the set of symbolic
expressions $\mathbb{E}$,
or $\top$ to represent no information: $v \in \{ e, \top \}, e \in \mathbb{E}$.
The mapping is updated through assignments, building more complex expressions
by substitution.
This domain allows reconstructing dynamically high-level
expressions from low-level stack-based evaluation, without requiring
a static pre-processing phase as done for instance by \cite{sawja}
on Java bytecode.
In our example, just before the \michelinstr{IF} instruction, the top of the
stack contains the expression \emph{eq(compare(x, y))}, which allows
\michelinstr{IF} to apply flow-sensitive constraints on
the \emph{x} and \emph{y} values.


\subsection{Equality Domain}

A stack-based execution model entails pushing copies of existing values
from the stack (using \michelinstr{DUP}), to be consumed
later by operators while leaving the original values intact for future use.
This can be seen, for instance, in \fref{fig:compare}.
In this context, maintaining information about variable equalities is
critical for precision: it allows any information inferred on one
copy to be propagated to other copies.
The symbolic expression domain from the last section helps to a degree, as it
allows substituting \texttt{x} with \texttt{y} after an assignment
\texttt{x := y}, which is sufficient for the case in \fref{fig:compare}.
However, this substitution mechanism is unidirectional and can fail
when the symmetry or the transitivity of equality is required.
Equalities can be tracked by numerical abstract domains, such as polyhedra,
but this is limited to numeric values, while we require tracking the
equality of values of complex types (such as maps, for the example
from \fref{fig:motiv}).
To solve this problem, we developed a simple domain able to infer
variable equalities.
It maintains a set of equivalence classes for variables that are
known to be equal. It proved to be more reliable than symbolic
expressions for the specific purpose of tracking equalities on non-numeric
variables.



\subsection{Symbolic Maps}\label{sec:smap}

On the example \fref{fig:motiv}, we want to prove that
only the owner of an account stored in the storage map can reduce its amount.
This requires inferring a numerical property about the contents of a map.
However, this is not a uniform property: the property on the value depends on
whether the key associated to it equals the \michelinstr{$sender} address or not.
Hence, the simple summarization abstraction of \sref{sec:types}
is not expressive enough.
We propose a map abstraction of the form:
$\{\mathit{sender}\mapsto\mathit{amount},
  \neg\mathit{sender}\mapsto\mathit{namount}\}$
that uses two variables per map:
$\mathit{amount}$ represents the value associated to the key equal
  to \michelinstr{$sender} for this call;
$\mathit{namount}$ summarizes all the values associated to other keys.

Like previous abstractions, $\mathit{amount}$ and $\mathit{namount}$
are variables, the values of which are abstracted in \michelinstr{mutez} domain.
Using a relational domain, it is possible to even track relations between
different versions and copies of the map from the storage.
All map operations are translated into operations on these variables,
depending on whether the key used to access the map equals \michelinstr{$sender}
or not, which can be precisely tested using our symbolic address domain
(\sref{sec:addresses}).

In our example, when updating the value of $\mathit{namount}$,
we check that the new value is greater than or equal to the previous one.
This is always the case for \emph{deposit} (even if the \emph{key} address
was specified in parameter), as transferred amounts are always non-negative.
As for \emph{withdraw}, updating the old value with
the value \texttt{(owned - asked)} triggers an error for the
original version without the fix.
For the fixed version, we are able to prove that the property is correct because the value of
$\mathit{namount}$ is unchanged in \emph{withdraw}.

\section{Experimental Results}
\label{sec:experiments}
\begin{table}[t]
\vspace{2mm}
\caption{Experimental evaluation}\label{table:exp}
\begin{tabular}{|r|r|r|r|r|r|r|r|r|r|}
        \hline
        Analysis                 & intv  & poly   & intv+exp & poly+exp \\
        \hline
        \hline
        total contracts               & 2931  & 2833   & 1579     &  1549    \\
        \hline
        mutez overflow           & 2824  & 1967   &  411     &  308     \\
        \hline
        shift overflow           & 10    & 10     &   9      &  9       \\
        \hline
        always fail      & 32    & 33     &  32      &  33      \\
        \hline
        min. time       & 0.076s & 0.15s   &  0.17s    &  0.16s    \\
        \hline
        max. time       & 71.02s & 568.25s & 581.34s   &  590.86s  \\
        \hline
        avg. time       & 3.31s  & 29.8s   & 26.43s    &  21.89s   \\
        \hline
\end{tabular}
\end{table}

We performed two kinds of experiments.
Firstly, we analyzed a large set of existing contracts for non-functional correctness
(e.g., absence of overflows) to assess the practicality and scalability
of our method.
Secondly, we analyzed more specifically the example from \fref{fig:motiv} for
our functional specification: only the owner of an account can decrease its amount.
We used a Xeon E5-2650 CPU with 128GB memory.
Our prototype can be found at \url{https://gitlab.com/baugr/mopsa-analyzer}
at commit tag \texttt{soap22}.

We selected the Carthagenet test network containing 2935 contracts with size
ranging from 1 to 3604 lines, and analyzed them with arbitrary storage.
The results, using different domain combinations, is presented in
\tref{table:exp}:
\emph{intv} uses the domains from \sref{sec:simpleanalysis} and the interval domain;
\emph{poly} adds the polyhedra domain;
\emph{intv+exp} and \emph{poly+exp} add the domains from \sref{sec:higher}.
The first line indicates the number of successful analyses (not all domains can support
all contracts due to the prototype nature of our implementation).
The lines \emph{mutez overflow} and \emph{shift overflow} indicate the number of runtime errors detected, whereas \emph{always fail} is the number of contracts always terminating in a failing state.
The last three lines indicate the minimal, maximal and average runtime per contract.
We expect that a large number of \emph{mutez overflows} are actually false positive
as the analysis assumes that arbitrary 63-bit amounts can be stored
and transferred but, in fact, the total number of \micheltype{mutez}
in circulation is far smaller.

Our prototype can check the functional correctness of our motivating
example from \fref{fig:motiv} in 0.273s.
As for other examples, it raises spurious overflows in \micheltype{mutez}
computations.

\section{Conclusion}
\label{sec:conclusion}
We have proposed a new sound and efficient static analysis based on
Abstract Interpretation for the Michelson smart-contract language.
Our prototype implemented in MOPSA is already able to analyze realistic
smart-contracts for runtime errors, and higher-level functional properties for
toy contracts using realistic authentication patterns.

Future work include strengthening our implementation to analyze more contracts,
as well as our support of operation lists for inter-contract analysis.
We will also focus on analyzing functional correctness properties, closing the gap between
our simplified example and the actual Dexter implementation, and considering
other smart-contracts and properties.
This entails developing more expressive domains, e.g. extending our non-uniform
map abstraction to arbitrary value types,
and supporting more complex authentication patterns, such as
cryptographic signatures.
Finally, we plan to exploit the value analysis to perform a gas
consumption (\ie timing) analysis.







\newpage
\bibliographystyle{ACM-Reference-Format}
\bibliography{bibliography}

\end{document}